\documentclass[preprint,floats,prl,aps]{revtex4}

\usepackage[dvips]{graphicx}
\usepackage{amsfonts}
\usepackage{amsmath}
\usepackage{amssymb}
\usepackage{exscale}

\begin{document}
\title{
	   Fermi surface instability at the hidden-order transition of URu$_{2}$Si$_{2}$.
	   }
\author{A. F. Santander-Syro$^{1,2\star}$, 
		M. Klein$^{3}$,
		F. L. Boariu$^{3}$, 
		A. Nuber$^{3}$,
		P. Lejay$^{4}$ 
		\& F. Reinert$^{3,5}$ 
		}
\address{$^1$Laboratoire Photons Et Mati\`ere, UPR-5 CNRS, ESPCI,
 	   	 10 rue Vauquelin, 75231 Paris cedex 5, France.}
\address{$^2$Laboratoire de Physique des Solides, UMR-8502 CNRS, Universit\'e Paris-Sud, 
		 91405 Orsay, France.}
\address{$^3$Universit\"at W\"urzburg, Experimentelle Physik II, Am Hubland, D-97074 W\"urzburg, 
	     Germany.}
\address{$^4$Institut N\'eel, CNRS/UJF, B.P. 166, 38042 Grenoble Cedex 9, France.}
\address{$^5$Forschungzentrum Karlsruhe, Gemeinschaftslabor f\"ur Nanoanalytik,
		 D-76021 Karlsruhe, Germany.}
\maketitle

{\bf 
Solids with strong electron correlations 
generally develop exotic phases of electron matter 
at low temperatures~\cite{vanHarlingen-dWave, MackenzieMaeno-Sr2RuO4, Dagotto-1Dto2DLadders,
Hotta-OrbitalOrdering-d-f-electrons, Mostovoy-HelicoidalOrdering-SrFeO3}. 
Among such systems, the heavy-fermion semi-metal URu$_{\bf 2}$Si$_{\bf 2}$ 
presents an enigmatic transition 
at {\it T$_{\bf o}$}~=~17.5~K to a `hidden order' state whose order parameter remains unknown 
after 23 years of intense research~\cite{Palstra-WhenItAllBegan-1985, Coleman-NV-Wiebe}. 
Various experiments point to the reconstruction and partial gapping of 
the Fermi surface when the hidden-order
establishes~\cite{Maple-Cv-PartiallyGappedFS, Ohkuni-dHvA, 
Palstra-AnisotropicResistivity-1986, Kamran-ThermoTransportURS, Wiebe-Neutrons-NatPhys,
Schoenes-HallEffectURS-1987, Klaus-PointContactURS-GapHO, Escudero-STS-HO-URS,
Bonn-IR-URS, Broholm-FirstNeutronData, Villaume-Neutrons}.
However, up to now, the question of how this transition affects the electronic spectrum 
at the Fermi surface has not been directly addressed by a spectroscopic probe.
Here we show, using angle-resolved photoemission spectroscopy, 
that a band of heavy quasi-particles drops below the Fermi level 
upon the transition to the hidden-order state. Our data provide 
the first direct evidence of a large reorganization of the electronic structure 
across the Fermi surface of URu$_{2}$Si$_{2}$ occurring during this transition, 
and unveil a new kind of Fermi-surface instability in correlated electron systems.
}

Earlier angle-resolved photoemission spectroscopy (ARPES) 
experiments mapped the basic band structure 
of URu$_2$Si$_2$ in the paramagnetic state (above $T_o$), 
establishing the existence of hole-pockets at the $\Gamma$, $Z$ and $X$ points 
of the Brillouin zone~\cite{Ito-ARPES-URS-30K, Denlinger-advancesARPES-f-electrons, 
Denlinger-ComparativeXRu2Si2}.
These experiments revealed strong disagreements with the calculations
for the electronic structure and Fermi surface
of URu$_2$Si$_2$. It was speculated that this was due to the
presence of narrow features from the U-5$f$ states,
not taken into account by the calculations, and
difficult to characterize experimentally with the resolutions
available at the time~\cite{Denlinger-ComparativeXRu2Si2}.
To date, no reports exist of high-resolution ARPES experiments
below or across $T_o$. The pressing question is to determine experimentally
the electronic structure near the Fermi level ($E_{F}$), inlcuding the heavy 5$f$ states,
above and below $T_o$.

\begin{figure}
   \begin{center}
      \includegraphics[width=12cm]{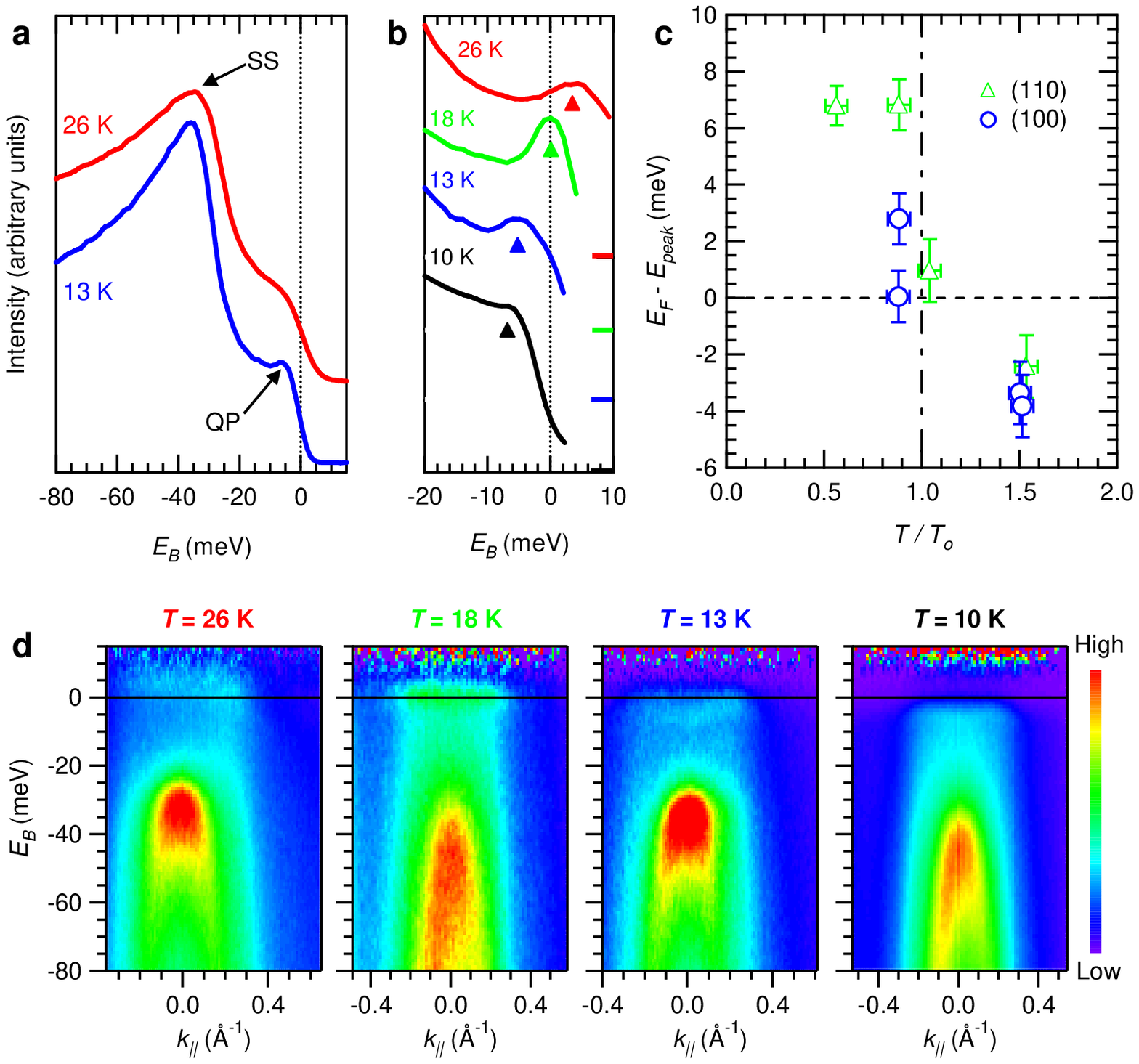}
   \end{center}
  \caption{\label{Fig1}
		   {\bf Temperature dependence of the quasi-particle band in URu$_{2}$Si$_{2}$}.
		   {\bf a}, Raw photoemission spectra integrated within 
		   $\pm 0.2$ \AA$^{-1}$ along the $(110)$ direction at 13~K
		   (blue) and 26~K (red). Below $T_o = 17.5$~K, 
		   a quasi-particle (QP) peak appears below $E_{F}$.
		   For all temperatures, a surface state (SS) at $E_{B} < -35$~meV is observed.
		   {\bf b}, Spectra integrated within $\pm 0.2$ \AA$^{-1}$ 
		   along the $(110)$ direction and normalized by the Fermi-Dirac distribution, at 
		   various temperatures around $T_o$: 26~K (red), 18~K (green),
		   13~K (blue) and 10~K (black). The zero-intensity level of each spectrum
		   is indicated by the color bars in the right axis.  
		   The triangle markers give the peak position.
		   {\bf c}, Energy of the QP peak in the integrated spectra 
		   (with respect to $E_{F}$) as a function of $T/T_o$ for cuts along
		   the $(110)$ (triangles) and $(100)$ (circles) directions. 
		   The error bars in $T$ are due to thermal instabilities
		   during the experiment. The error bars in the peak energy
		   are calculated from the peak positions in spectra integrated over different 
		   momenta windows around $k_{\parallel} = 0$.
		   {\bf d}, Angle-resolved spectra along the $(110)$ direction for the
		   same temperatures as in {\bf b}, over an extended energy range. 
		   The itense hole-like feature dispersing below $E_B \sim -35$~meV 
		   is the surface state displayed in panel {\bf a}.
		   }
\end{figure}

Figure~\ref{Fig1} summarizes our findings for the temperature evolution
of the electronic structure near $E_F$. 
Figure~\ref{Fig1}a shows the angle-integrated spectra of electrons 
with $k_{\parallel}$, the momentum component parallel to the sample surface,
along the $(110)$ direction 
at two temperatures across the transition.  
At $T = 26$~K, the only apparent feature is a 
surface state (SS) at binding energies $E_{B} < -35$~meV, 
observed at all the investigated temperatures 
(see Supplementary Material).
In contrast, at 13~K a narrow peak at $E_B ~\approx -7$~meV appears,
signaling the presence of a quasi-particle (QP) band. The temperature
dependence of this QP band was systematically studied,  
and is shown in Figures~\ref{Fig1}b-d. 
In these figures we normalized the spectra by the Fermi-Dirac distribution,
following a well established procedure~\cite{Ehm-KondoResonance-FDDdivision},
to reveal the thermally occupied part of the spectral function up to
energies $\sim 5k_{B}T$ above $E_F$. 
The angle-integrated data of Fig.~\ref{Fig1}b shows that at 26~K the QP band
lies at $E_B \approx 5$~meV,  at 18~K~$\approx T_o$ it appears right at $E_{F}$, and
below $T_o$ the band shifts to energies below $E_F$. At 10~K the QP 
peak is located at $E_B \approx -7$~meV.
Figure~\ref{Fig1}c presents a quantitative evaluation of the QP peak energy 
as a function of $T/T_o$ from several spectra taken along
the $(110)$ and $(100)$ directions, 
implying that the shift of this QP band occurs 
over an extended region of momentum space.
Figure~\ref{Fig1}d displays the angle-resolved data at the same temperatures as
in Fig.~\ref{Fig1}b. 
The data at 26~K and 18~K show, respectively, 
a flat band (within resolution) above $E_F$ and at $E_F$.
These correspond to the peaks observed in the angle-integrated data of Fig.~\ref{Fig1}b 
at those temperatures. Interestingly, the spectral weight of this flat band 
appears to be confined to a momentum region within
$k_{\parallel}=\pm 0.2$~\AA$^{-1}$, momenta at which there is a clear hint of 
a Fermi-level crossing. 
In the spectra at 13~K and 10~K, {\it i.e.} below $T_o$, 
the QP band displays a narrow dispersion, 
showing that it is a band of 
itinerant heavy quasi-particles. 
We will discuss all these observations in detail. 

The data of Figure~\ref{Fig1} show explicitly 
that the transition to the HO state is accompanied by a significant reorganization 
of the portions of the Fermi surface involving heavy quasi-particles.
Such a transfer of spectral weight across the Fermi surface
implies dramatic modifications of the macroscopic properties of the system,
strongly suggesting that this is the microscopic origin of the observed abrupt changes
in the thermal and transport properties of URu$_2$Si$_2$ 
during the HO transition~\cite{Palstra-WhenItAllBegan-1985, 
Maple-Cv-PartiallyGappedFS, Palstra-AnisotropicResistivity-1986, 
Kamran-ThermoTransportURS, Schoenes-HallEffectURS-1987}.

\begin{figure}
   \begin{center}
      \includegraphics{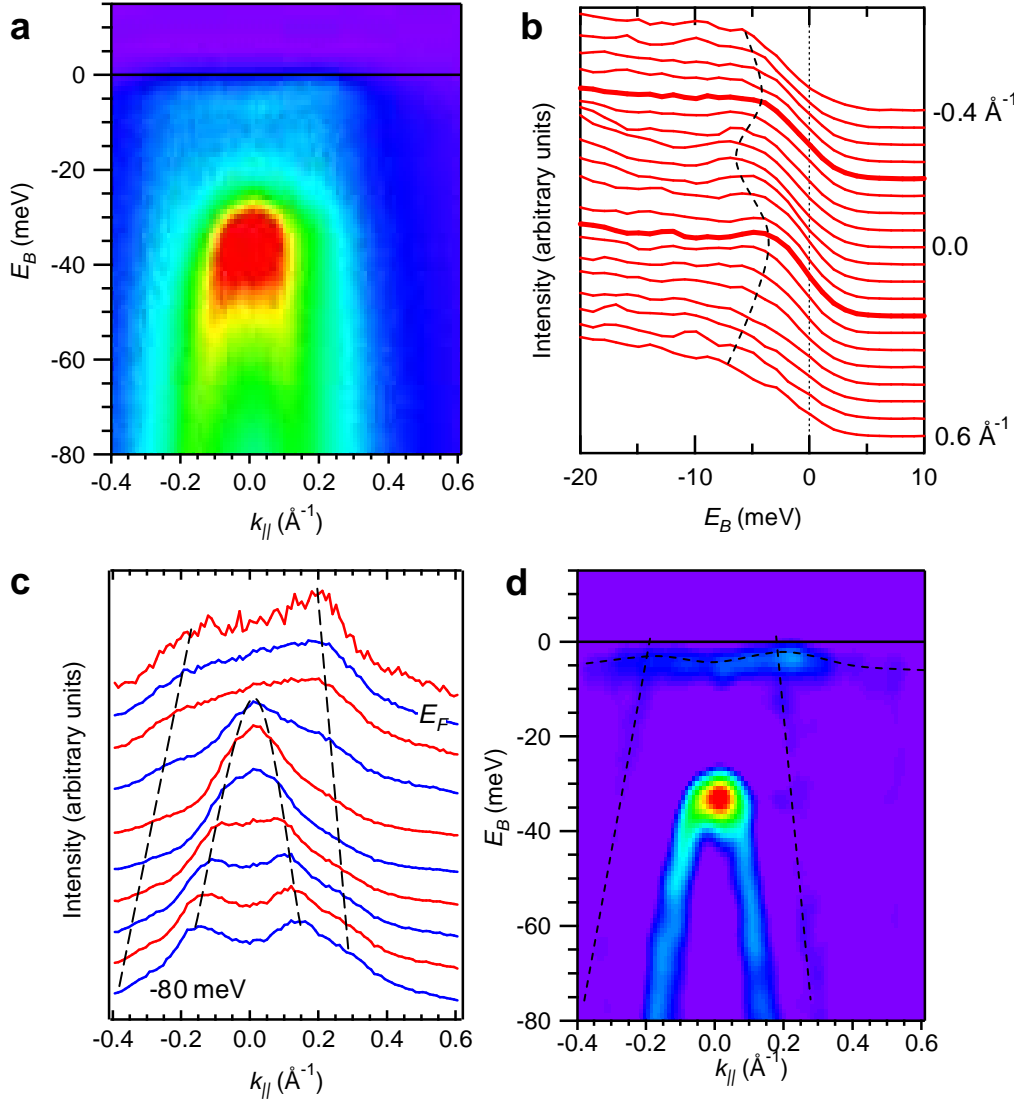}
   \end{center}
  \caption{\label{Fig2}
		   {\bf Heavy quasi-particle band in the hidden-order state and hybridisation 
		   with a conduction band along the $(110)$ direction}.
		   {\bf a},~ARPES intensity map along the $(110)$ direction
		   at 13~K.  The map shows a heavy 
		   quasi-particle band dispersing down to $\sim 7$~meV below $E_{F}$.
		   {\bf b},~EDCs of the intensity map in ({\bf a}) in the region close to $E_{F}$. 
		   The EDCs whose leading edge is closest to $E_F$ are drawn in bold. 
		   {\bf c},~MDCs from the intensity map in ({\bf a}). Each MDC is normalized to its
		   area. The two central peaks correspond to the hole-like surface state,
		   and two lateral shoulders to a light 
		   conduction band that disperses through $E_{F}$.
		   {\bf d},~Average of second derivatives along the energy and momentum axes
		   (see methods), showing the heavy-QP band, the surface state, and the hole-like
		   conduction band.
		   In all panels, the dashed lines are guides to the eye for the dispersions
		   of the different bands.
		   }
\end{figure}

\begin{figure}
   \begin{center}
      \includegraphics{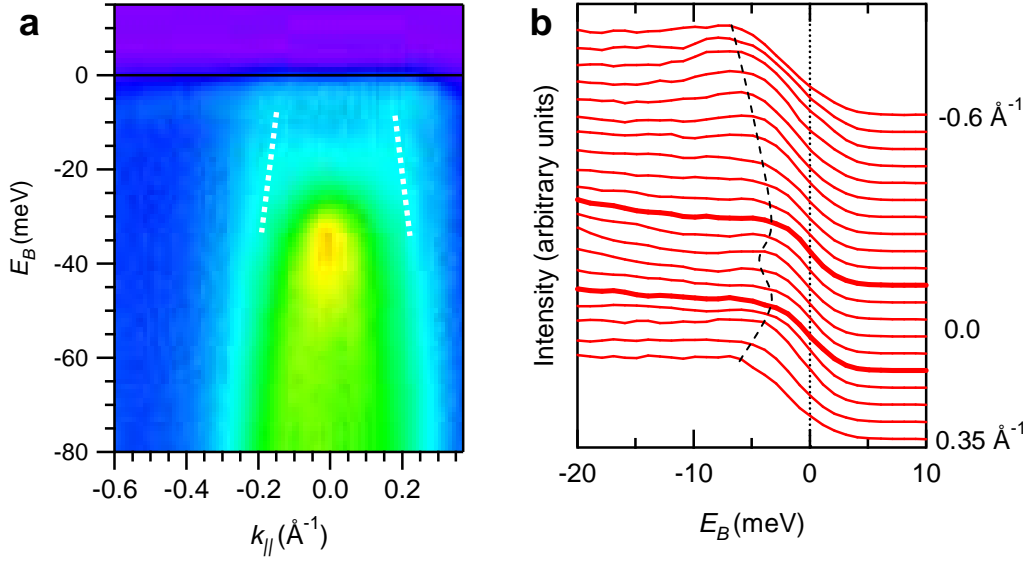}
   \end{center}
  \caption{\label{Fig3}
		   {\bf Heavy quasi-particle band in the hidden-order state and hybridisation 
		   with a conduction band along the $(100)$ direction}.
		   {\bf a},~ARPES intensity map along the $(100)$ direction
		   at 15~K, with the heavy-QP band dispersing down to $\sim 4$~meV below $E_{F}$.
		   The dashed white lines show conduction bands dispersing through $E_{F}$,
		   clearly visible in the raw map.
		   {\bf b},~EDCs of the intensity map in ({\bf a}) in the region close to $E_{F}$. 
		   The EDCs whose leading edge is closest to $E_F$ are drawn in bold.
		   As a guide to the eye, the dispersion of the heavy-QP band 
		   is shown by the dashed line. This band, clearly extending 
		   beyond the Fermi momenta,
		   is also visible in the raw data map of panel {\bf a}.
		   }
\end{figure}

Figure~\ref{Fig2} shows the angle-resolved data at $T=13$~K, along the $(110)$ direction. 
The ARPES intensity map (Fig.~\ref{Fig2}a) shows 
the narrow quasi-particle band dispersing 
down to $W \approx -7$~meV, where $|W|$ is the width of the band.
Figure~\ref{Fig2}b shows energy distribution (EDCs) curves 
in the region close to $E_{F}$ of this intensity 
map. The EDCs having the leading edge (LE)
closest to $E_{F}$ are plotted in bold, corresponding to momenta 
$k_{LE} = \pm 0.2$~\AA$^{-1}$.
For $|k|<|k_{LE|}$ distinct QP peaks of resolution-limited width ($\sim 5$~meV)
are observed. 
From the values of $W$ and $k_{LE}$, an estimate of the effective mass 
($m^{\star}$) of these QPs can be obtained from
the relation $W = -\hbar^2 k_{LE}^{2}/2m^{\star}$.  This yields $m^{\star}\approx 22 m_e$
($m_e$ is the bare electron mass),
confirming that this band corresponds to heavy quasi-particles.  
This value of $m^{\star}$ is among the largest values  
ever measured by ARPES in any material. 
The values of $m^{\star}$ and $k_{LE}$ agree well with values given by
specific heat data~\cite{Maple-Cv-PartiallyGappedFS}
and de~Haas-van~Alphen measurements~\cite{Ohkuni-dHvA}.
The group velocity ($v_{QP}$) of the observed heavy-QPs can also be estimated
from $v_{QP} \approx W/k_{LE}$, yielding $|v_{QP}| \approx 35$~meV~\AA, comparable
to values obtained from thermal transport data~\cite{Kamran-ThermoTransportURS}.
Notice also from Fig.~\ref{Fig2}b that for $|k|>|k_{LE}|$ 
the leading edge of the spectra shifts to larger binding energies.
This suggests that the spectral weight of the heavy-QP band spreads
over a large momentum window, 
as best seen in data along the $(100)$ direction (see Fig.\ref{Fig3}). 
Figure~\ref{Fig2}c shows momentum distribution curves (MDCs) 
from the intensity map in Fig.~\ref{Fig2}a.
Besides the peaks corresponding to the hole-like surface state,
two lateral shoulders are observed (shown by the dashed lines). 
They correspond to a light hole-like conduction band 
with $m^{\star} \approx -1.4 m_e$, dispersing through $E_{F}$
at Fermi momenta $k_F = \pm 0.2$~\AA$^{-1}$, 
{\it i.e.}, $k_F = k_{LE}$ within experimental resolution.
Fig.~\ref{Fig2}d shows the average of the energy and momentum second derivatives 
of the intensity map in Fig.~\ref{Fig2}a. This allows to visualize clearly the 
conduction band and the spectral weight of the heavy-QP band spreading beyond $|k_{LE}|$.

Figure~\ref{Fig3} summarizes the data along the $(100)$ direction at $T<T_o$.
The raw map (Fig.~\ref{Fig3}a) clearly displays 
the band of itinerant heavy QPs approaching $E_F$ at $k_{LE} = \pm 0.15$~\AA$^{-1}$,
the light hole-like conduction band dispersing through $E_{F}$ at the same momenta,
and the tails of spectral weight extending beyond $|k_{LE}|$.
The latter have a clear quasi-particle peak structure, 
with a hole-like dispersion of velocity $\sim 12$~meV~\AA, as seen in the
corresponding EDCs (Fig.~\ref{Fig3}b).

An interesting observation from Figs.~\ref{Fig2} and~\ref{Fig3} 
is that the momenta where the conduction band and heavy-QP coincide 
correspond to the momenta where the heavy-QP band bends back from $E_F$. 
This suggests that the whole structure arises from the hybridization 
of the light conduction band and a band of localized states, 
though due to the finite experimental resolution, 
we cannot observe a hybridization gap directly.
From our angle-resolved data at 26~K and 18~K in Fig.~\ref{Fig1}d,
it appears that these these bands hybridize already {\it above} $T_o$, when the 
band of localized states is at $E_B>E_F$. 
The transition to the hidden-order state shifts 
the heavy-QP part of the resulting spectral function to $E_B < E_F$. There,
it is clearly observed as a dispersing band of heavy quasi-particles, 
creating --together with the light conduction band-- a heavy Fermi surface. 
Indeed, previous ARPES experiments in URu$_2$Si$_2$ above $T_o$ 
showed strong evidence for the presence of 
$f-d$ hybridization~\cite{Denlinger-ComparativeXRu2Si2}. 

Note also that along the $(100)$ direction $k_F$ is lower than along the $(110)$ direction.
This provides direct experimental evidence that anisotropic small-sized 
Fermi-surface pockets exist around the $\Gamma$ point in URu$_2$Si$_2$,
as indirectly suggested by other techniques~\cite{Ohkuni-dHvA}.

Summarizing, our results explicitly show that the hidden-order transition of
URu$_2$Si$_2$ results in a large transfer of spectral weight across
the Fermi surface. Below the ordering temperature, 
our data reveal the existence of a band of heavy quasi-particles
dispersing over a narrow energy scale of the order of $7$~meV
below $E_F$.
The observations suggest that this heavy-QP band arises from the
hybridization of a $d$-conduction band with a 
band of localized states, 
probably of $5f$ character~\cite{Denlinger-ComparativeXRu2Si2}.
As temperature rises above $T_o$, the heavy-QP band 
moves to unoccupied states above $E_F$.
These results, which cannot be understood in terms of 
band-nesting alone, demonstrate that the interplay of localized-itinerant behaviors of
the electrons in URu$_2$Si$_2$ is important to understand the hidden-order transition.
Moreover, these findings 
emphasize the fundamental role of heavy quasi-particles  
during the transition, as it is their spectral weight that shifts 
to below $E_F$ when the ordered state sets in. 
This remarkable phenomenon, which is in itself 
a novel kind of Fermi-surface instability in solids,
has to be taken into account in theories of the hidden-order transition 
in URu$_2$Si$_2$.

One way of interpreting our data is that the hidden-order 
transition massively re-organizes the spectral function of this 
strongly-interacting many-electron system.  The observed spectral-weight
shift of the heavy quasi-particle band would then be a consequence of this many-body effect.
A proof of principle of this possibility, that does not need to invoke band-nesting,
has been recently given for the case of the magnetic order-disorder
transition in the two-dimensional doped Kondo-lattice model~\cite{Fakhed}.
As the small magnetic moment observed in the HO state of URu$_2$Si$_2$ seems to have 
an extrinsic origin~\cite{Amitsuka-PT-phaseDiagram}
and appears anyway insufficient to explain the large entropy loss 
due to the transition~\cite{Wiebe-Neutrons-NatPhys},
material-specific calculations for the case of URu$_2$Si$_2$ would be needed 
to test this idea.

We anticipate that our results will be an invigorating trigger for more, 
high-resolution studies of the angular-resolved electronic structure of URu$_2$Si$_2$,
and for theoretical developments exploring how the competition between itinerant and
localized electron behavior 
can describe the fascinating behaviour of this material.
Our observations will provide new insight not only to the theoretical approaches 
of the hidden-order, but also to the understanding of exotic
phases in systems with strong electron interactions and competing ground states.

\section{Methods}
\subsection{Sample preparation and measurement technique}
The high-quality URu$_{2}$Si$_{2}$ single crystals were grown
in a tri-arc furnace equipped with a Czochralski puller,
and subsequently annealed at 900$^{\circ}$C under ultra high vacuum 
for 10 days~\cite{Lejay-XtalGrowth}.  
The high-resolution ARPES experiments were performed
with a Gammadata R4000 analyzer and a
monochromatized VUV-lamp at $h\nu = 21.2$~eV (He~I$_{\alpha}$).
The energy resolution for the used analyzer settings
was 5.18~meV, determined similar to previous experiments~\cite{Reinert-BCSV3Si}.
The temperature during each measurement was found by fitting
a Fermi-Dirac distribution to the Fermi edge of
polycrystalline Ag, mounted next to each URu$_{2}$Si$_{2}$
crystal on the same sample holder, and taking into account the above energy resolution. 
The base pressure in the chamber was $1\times10^{-10}$~mbar,
increasing to $8\times10^{-10}$~mbar during the measurement due
to the He leakage from the discharge lamp.
The crystals were oriented using Laue diffraction, and cleaved {\it in situ} 
just before the measurement, already at the measurement temperature. 
Highly ordered surfaces were confirmed by sharp low-energy electron diffraction
patterns measured on each sample after the measurements.
Because of the observed high-surface reactivity (the quasi-particle peaks below $T_o$
broaden and lose amplitude within 2 hours after cleaveage), 
the duration of each measurement was kept below 15 minutes immediatly
after cleaving, at constant temperature.
\subsection{Procedure of second-derivative rendering}
The raw photoemission intensity maps were convoluted with a two-dimensional gaussian
of widths $\sigma_{E}=7$~meV and $\sigma_{k}=0.07$~\AA$^{-1}$.  
Second derivatives along the $E_{B}$ and $k_{\parallel}$ axes were
normalized to the maximum intensity peak in the surface-state, then averaged.
Only negative intensity values are shown to trace the peaks of
the raw data. 

\section{Acknowledgements}
We thank F. Assaad, L. Bascones, K. Behnia and J. Flouquet
for illuminating discussions. A.F.S.S thanks LPEM for financial support. The work at 
University of W\"urzburg was supported by the Deutsche Forschungsgemeinschaft
through grant No. Re~1469/4-3/4 (M.K., F.L.B., A.N., F.R.)

\section{Competing interests} 
The authors declare that they have no
competing financial interests.

\section{Additional information} 
Correspondence and requests for materials
should be addressed to A.F.S.S. (e-mail:~andres.santander@espci.fr).



\newpage

\end{document}


\title{SUPPLEMENTARY MATERIAL\\
           Fermi surface instability at the hidden-order transition of URu$_{2}$Si$_{2}$
           }
\author{A. F. Santander-Syro$^{1,2}$, 
        M. Klein$^{3}$,
        F. L. Boariu$^{3}$, 
        A. Nuber$^{3}$,
		P. Lejay$^{4}$ 
		\& F. Reinert$^{3,5}$ 
        }
\address{$^1$Laboratoire Photons Et Mati\`ere, UPR-5 CNRS, ESPCI, 10 rue Vauquelin, 
                75231 Paris cedex 5, France}
\address{$^2$Laboratoire de Physique des Solides, UMR-8502 CNRS, Universit\'e Paris-Sud, 
                91405 Orsay, France}
\address{$^3$Universit\"at W\"urzburg, Experimentelle Physik II, Am Hubland, D-97074 W\"urzburg, 
                Germany}
\address{$^4$Institut N\'eel, CNRS/UJF, B.P. 166, 38042 Grenoble Cedex 9, France}
\address{$^5$Forschungzentrum Karlsruhe, Gemeinschaftslabor f\"ur Nanoanalytik,
                D-76021 Karlsruhe, Germany}
\begin{abstract}
The photoemission data of URu$_2$Si$_2$ show an intense hole-like feature with a minimal 
binding energy of $E_0\approx-35$~meV right after cleaving the crystal {\it in situ}. 
We briefly present here the evidence that it is a surface state.
\end{abstract}
\maketitle

\begin{figure}
        \begin{center}
                \includegraphics[width=8cm]{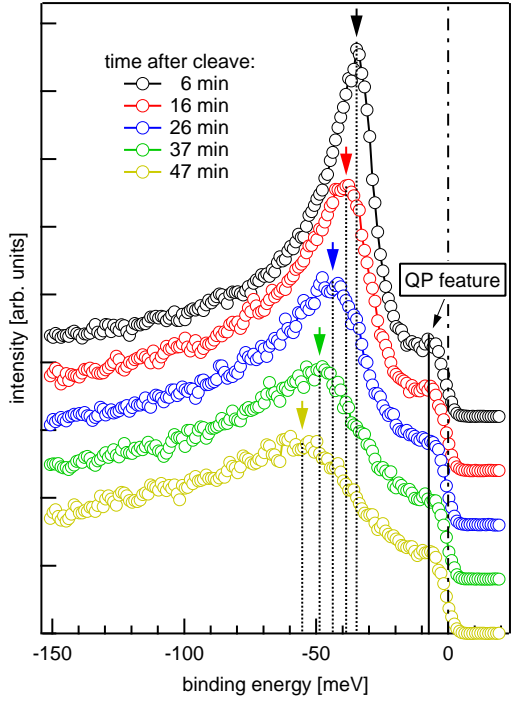}
    \end{center}
        \caption{\label{ads}
                        EDCs at normal emission obtained 
                        at different times after cleaving the crystal. 
                        The intense feature's energy shifts with time due to adsorption 
                        of residual gases in the UHV chamber, characteristic of a
                        surface state. The heavy quasi-particle (QP) feature right below $E_F$ 
                        remains at constant energy, indicative of a bulk feature.
                        }
\end{figure}

\begin{figure}
        \begin{center}
        \includegraphics[width=8cm]{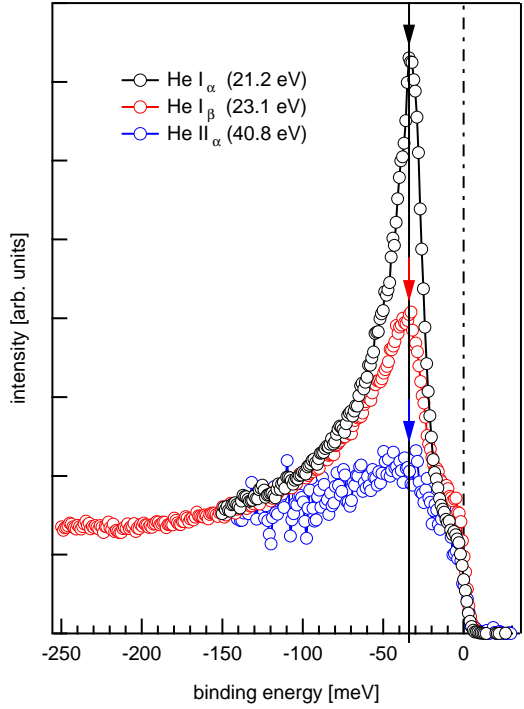}
        \end{center}
        \caption{\label{hv}
                        EDCs at normal emission measured with different excitation energies. 
                        The marked feature stays at a constant energy position showing 
                        no dispersion in $k_z$. This indicates a two dimensional state,
                        which in our case is a surface state.}
\end{figure}

Measuring at different times after the cleavage results in an energy shift 
of such feature. Figure~SF~\ref{ads} shows a measurement series with 
energy distribution curves (EDCs) at normal emission obtained in time intervals 
of about 10~min. The peak positions are marked by arrows. This energy shift with time is due 
to the adsorption by the sample's surface of residual gas, mainly originating from 
leakage of the He discharge lamp,
to which the states located at the surface are highly sensitive.
A similar behavior of noble metal surface states has been observed 
by Nicolay and coworkers~\cite{Nicolay2003}. By contrast, note that 
the heavy-quasi-particle (QP) feature 
right below $E_F$ remains at a constant energy position, 
indicating a bulk-like character.

Figure~SF~\ref{hv} shows EDCs at normal emission measured with different excitation energies
{\it right after cleaving}. 
In this case the position in energy of the surface related feature is constant. 
This means 
that such a state shows no dispersion in the $k_z$ direction 
and hence has a two-dimensional character. This fully confirms that the intense feature 
below $E_0\approx-35$~meV is a surface state. 
%


%